\documentclass[11pt]{revtex4}
\usepackage{graphicx}
\usepackage{amssymb}
\usepackage{amsmath, amsfonts}
\usepackage{epstopdf}
\usepackage{subfigure}
\usepackage{dcolumn}
\usepackage[usenames,dvipsnames]{color}   
\usepackage{parskip}
\usepackage{hyperref}
\usepackage{placeins}
\usepackage[font=small,labelfont=bf]{caption}
\usepackage{sidecap}
\usepackage{threeparttable}
\usepackage{ifthen}

\begin{document}

\title{
Impact of vibrational entropy on the stability of unsolvated peptide helices with increasing length}

\author{Mariana Rossi, Volker Blum, and Matthias Scheffler}
\affiliation{Fritz-Haber-Institut der Max-Planck-Gesellschaft, Faradayweg 4-6, 14195 Berlin, Germany}

\begin{abstract} 
Helices are a key folding motif in protein structure.
The question which factors determine helix stability for a given 
polypeptide or protein is an ongoing challenge.
Here we use van der Waals corrected density-functional
theory to address a part of this question in a bottom-up approach.
We show how intrinsic helical structure is stabilized with
length and temperature for a series of experimentally well studied 
unsolvated alanine based polypeptides, Ac-Ala$_n$-LysH$^+$.
By exploring extensively the conformational
space of these molecules, we find that helices emerge as the preferred structure in the length range
 $n$=4-8 not just due to enthalpic factors (hydrogen bonds and their cooperativity,
van der Waals dispersion interactions, electrostatics), but importantly also by a vibrational entropic stabilization
over competing conformers at room temperature.
The stabilization is shown to be due to softer low-frequency
vibrational modes in helical conformers than in more compact ones.
This observation is corroborated by including anharmonic effects explicitly through \emph{ab initio} molecular dynamics,
and generalized by testing different terminations and considering larger helical peptide models.  
\end{abstract}

\maketitle




\section{Introduction}

Polypeptide helices are a key secondary structure motif 
in a wide range of proteins \cite{Pauling1951_1, pdb, Creighton1987}.
It is well known that some amino acids (e.g., alanine) exhibit a
stronger helix propensity than others \cite{PaceScholtz1998,ScholtzBaldwin1991,HorovitzFersht1992,CreamerRose1992,Kinnear:2001p4453, MillerKemp2002,ScottFersht2007, Moreau2009}, 
but the fact that the helical structure 
is so abundant in proteins is still intriguing. From a thermodynamic
point of view, 
there are at least two possible limits in which helices compete
with other structure prototypes. Towards high temperature,
one expects the transition to a random coil \cite{ZimmBragg1959},
which should become entropically favored as the
temperature increases. Towards low temperature, however, 
helices may themselves compete with other enthalpically stable conformations.
However, in the most interesting regime, namely intermediate, physiological temperatures,
stability may be determined by a delicate balance between enthalpy and entropy \cite{Kinnear:2002p4458}. 
We here unravel this balance
quantitatively for the emergence of helical structure in a
particularly well studied series of unsolvated polyalanine based peptides
Ac-Ala$_n$-LysH$^+$, $n$=4-8 \cite{HudginsJarrold1998, KohtaniJarroldWater2004}. We consider explicitly
not just the helical part of conformational space, but actually the
much larger, general low-energy conformational space of the
peptides, of which helices are a part. In this paper we show: (i) a comprehensive
search of the conformational space for Ac-Ala$_n$-LysH$^+$, $n$=4-8; (ii) harmonic free energy
calculations for several structural candidates; (iii) outlook on the role of anharmonicities in the potential
energy surface; and (iv) a theoretical comparison for longer model
peptides, considering only helical motifs, with a different
termination, in order to clarify the impact of Lys on the soft
vibrational modes.  
Our key finding is that there is a significant
vibrational entropic stabilization of helices compared to other, more
compact conformers, a contribution that should indeed make a
difference in actual proteins as well. This contribution is intrinsic
to the helix and should therefore act largely independently, not
entangled with environment-dependent terms such as a solvent
entropy. Interestingly, an essential role of low-frequency modes is
also actively debated in other areas of protein science, related to
their function \cite{TournierSmith2003, 
  BrooksKarplus1985, NatChemFocus,   HayScrutton2012,
  GlowackiHarveyMulholland2012}.

Beginning with the terms that shape the potential-energy surface (PES), 
known reasons for helix stability include \cite{ScholtzBaldwin1992,Baldwin-review}
(i) their efficient hydrogen-bond (H-bond) network and increasing H-bond
cooperativity with helix length \cite{GuoKarplus1994,Baldwin2003,IretaGalvan2003,dannenbergccoperativity2,TkatchenkoRossi2011},
(ii) suitably bonded and/or electrostatically favorable termination, for instance the LysH$^+$ termination considered here  
\cite{BlagdonGoodman1975,PrestaRose1988,SerranoFersht1989,TakahashiOoi1989,HudginsJarrold1998,MarquseeBaldwin1989,DugourdJarrold2005, WilliamsKemp1998},
and (iii) remarkably, a rather specific favorable contribution
of van der Waals (vdW) interactions for $\alpha$-helices \cite{TkatchenkoRossi2011,ShuguiShuhua2011}. 
Clearly, the peptide chain length plays a role: 
Too short
chains have too few and too weak hydrogen bonds for helices to
compensate the cost of strain in the
backbone \cite{MarquseeBaldwin1989,KohtaniJarroldWater2004,
  LiuBowers2004, JobKemp2006, 
  ElstnerSuhai2000}. 
In practice, environment
effects will necessarily influence 
helix stability \cite{ElstnerSuhai2000, SalvadorDannenberg2007,
  Garcia2002, TiradoRives-Jorgensen2000}. 
In an aqueous medium,  the hydrophobic effect will be prominent \cite{Kauzmann1959}. 
In water-poor conditions, like membranes, helices are frequently observed
\cite{MacKenzieEngelman1997,Vitaly2003}.
Finally, in vacuum conditions, the longer members ($n\ge$8) of the polyalanine based peptide
series studied here assume helical structure 
in experiment \cite{HudginsJarrold1998,JarroldReview2007,Rossi2010}.
These helices are stable \textit{in vacuo} even up to
extreme temperatures (not expected in solution) \cite{TkatchenkoRossi2011,KohtaniHighT:2004}, or
after soft landing on a surface \cite{WangLaskin2010}. 

The potential energy surface also shapes entropy, and
thus the effect of temperature $T$.
With increasing $T$, the conformational entropy of the backbone will favor an unfolded
state \cite{ZimmBragg1959,Baldwin-review, BrooksKarplusPettitt, PoulainCalvoDugourd2007} 
(so-called ``random coil''), while at low $T$ helices may also
compete with other, enthalpically more stable conformers. 
For instance, gas-phase ion mobility spectrometry (IMS) by
Jarrold and coworkers \cite{Kinnear:2002p4458} showed that the
Ac-Ala$_4$-Gly$_7$-Ala$_4$H$^+$ polypeptide is helical at $T$=400 K but
globular at room temperature. A similar structural change was observed in experiments involving 
multiply protonated polyalanine in the gas-phase in Ref. \cite{CountermanClemmer2003}.
Empirical force-field based simulations by Ma and coworkers  \cite{MaTsaiNussinov2000} of
more than 60 small peptides indicate that the \emph{vibrational}
entropy (harmonic approximation) could stabilize $\alpha$-helices or 
$\beta$-hairpins over competing low-temperature conformers.
Very recently, Plowright and coworkers \cite{PlowrightMons2011}
used density-functional theory (DFT) including dispersion contributions
(the B97-D \cite{Grimme2006} exchange-correlation functional) to suggest that,
for a small neutral four-residue peptide, 
$\beta$-sheets and conformers containing 3$_{10}$ helical loops
are stabilized by the harmonic
vibrational entropy at finite temperatures.

In the present work, we provide independent, unambiguous, and
quantitative computational evidence that the vibrational entropy acts
to stabilize helical conformers with increasing temperature over more
compact, enthalpically competitive structures. 
The reason, in short, is
traced to the softer low-frequency modes of helices, which are also reflected
in the dynamics (anharmonic case). We focus on
polyalanine-based peptides, 
since alanine is known to have a high helix propensity both 
in solution \cite{ScholtzBaldwin1992, ChakBaldwin1995} and \textit{in
vacuo} \cite{Kinnear:2001p4453}. For Ac-Ala$_n$-LysH$^+$ ($n$=4-20) in
the gas phase, IMS \cite{KohtaniJarroldWater2004} and first-principles
calculations compared to experimental vibrational spectroscopy at room
temperature \cite{Rossi2010} suggest a cross-over from non-helical to
helical preferred conformers as a function of polyalanine chain
length.  
For $n$=5 there is a competition between different conformers, while
the $n$=10 and $n$=15  conformers are found to be firmly in the
helical range \cite{Rossi2010}. 
In a previous publication \cite{TkatchenkoRossi2011} we have 
quantified, from first principles, the contributions from electrostatics, H-bond cooperativity,
and van der Waals interactions on the stability of unsolvated
polylalanine-based helices against unfolding.
This class of systems is thus an ideal testing
ground to clarify the structural competition of non-helical (compact) and helical
conformers as a function of chain length also toward the opposite
temperature limit (low temperature, folded state). 
In the following, we address conformational preference of Ac-Ala$_n$-LysH$^+$
\emph{in vacuo} for $n$=4-8, i.e., the length range in which the
helical preference at room temperature develops. 

Our 
work is based on an exhaustive prediction of low-energy conformers using
DFT and the PBE 
\cite{PBE} exchange-correlation functional, corrected to account for
 long-range vdW interactions
\cite{tkatchenko-scheffler2009} (here called PBE+vdW). This level of
theory treats accurately, and without system-specific empirical parameters,
critical length-dependent contributions 
such as H-bond cooperativity \cite{GuoKarplus1994,IretaGalvan2003,
  dannenbergccoperativity2,TkatchenkoRossi2011} and 
vdW interactions \cite{TkatchenkoRossi2011}, including their
effect on vibrational frequencies and in \emph{ab initio}
molecular dynamics.

\section{Methods}

Our conformational search strategy, used to find structure candidates of Ac-Ala$_n$LysH$^+$, $n$=4-8, 
consists of two steps. For a detailed
discussion of the search strategy we use here, we refer the reader to Ref. \cite{mythesis}. 
Both the Lys residue and the C-terminal COOH are considered protonated
throughout, as is known to be the gas-phase preference when the 
N-terminus is capped.\cite{HudginsJarrold1998, mclean2010, WieczorekDannenberg2004}

In the first step, we begin by an extensive, unconstrained 
force field (OPLS-AA
\cite{KaminskiJorgensenoplsaa2001}) basin-hopping search (using the
TINKER package \cite{tinker}), 
aiming to simply ``list'' as many different
conformers as possible. The same strategy was employed in
Ref. \cite{Rossi2010}, enumerating a huge number of candidate 
structures: at least 10$^5$ conformers for each of the molecules in
question. Our particular choice of the OPLS-AA force field is not
motivated by any other reason than that an input structure
``generator'' for DFT was needed. In particular, care was taken that
small changes to the parameters in the OPLS-AA part of the search
do not affect the structure manifold considered in the DFT step
below.\cite{mythesis} 

In the second step, a wide range of conformers suggested by the force field is fully relaxed 
using DFT with the PBE+vdW exchange-correlation functional in all-electron total energy calculations (FHI-aims
program package \cite{aims-Blum:2009}). We employ 
essentially converged numeric atom-centered basis sets and other
numerical settings \cite{aims-Blum:2009,Havu2009} for the
relaxations.
In detail, the PBE+vdW part of our searches
covers 1068, 1000, 800, 800, and 820 conformers for $n$=4, 5, 6, 7, 8, 
respectively. For $n$=6, 7, and 8, these numbers are chosen to ensure that at least
all conformers identified in the lowest 7 kcal/mol ($\approx$ 0.3 eV)
of the force field step are included in the PBE+vdW relaxations. 
For $n$=4 and 5, we explored the limits of our search
strategy \cite{mythesis}. When comparing locally relaxed minima in
PBE+vdW that were obtained by starting from an OPLS-AA local PES
minimum structure, we observe a correlation of the relative energy
hierarchies, but with a large scatter (max. around 50 meV per
residue). Therefore, a large number of conformers must be considered for
post-relaxation in PBE+vdW. The comparison also reveals structure-specific
force field errors that might otherwise go unnoticed. For example, we observe that
OPLS-AA systematically overestimates the energy of 3$_{10}$-helical
structures \emph{in vacuo}. 
To exclude any unwanted impact of the overestimation of 3$_{10}$-helices, we
performed additional constrained basin-hopping searches in which we
forced certain H-bonds of the molecule to remain 3$_{10}$-helical, again
followed by individual, unconstrained PBE+vdW relaxations.

The PBE+vdW relaxed conformers were sorted into
``families'' according to their H-bond pattern. We define an H-bond
to be present when an O acceptor atom is closer than 2.5 {\AA} to a
H donor atom. Within each H-bond family thus defined, small
conformational variations are still possible, e.g., by slight bends of
the backbone atoms or different rotamers of the LysH$^+$ side chain. 
Typically, the lowest-energy PBE+vdW conformer is found among the 
family members arising within 5 kcal/mol (0.2 eV) of the lowest-energy force field
conformer. 

Harmonic vibrational frequencies and intensities were computed from finite differences. 
The accuracy of the vibrational frequencies calculated in this manner is estimated by analyzing the rotational and translational modes. 
These modes should be at zero frequency, and thus the deviation observed gives a limit for the accuracy of the rest of the frequencies. 
We have not observed deviations of more than 2 cm$^{-1}$ in any calculations.
Harmonic free energies were calculated in the harmonic oscillator/rigid body approximation. 
Vibrational density of states beyond the harmonic approximation were calculated from the Fourier transform
of the velocity time autocorrelation function, taken from \textit{ab initio} molecular dynamics trajectories.

For all molecules containing $n$>8 alanine residues and for the Li$^+$
terminated model structures discussed in this work, no extensive
conformational search was performed. These peptides are structure
models used specifically for a computer experiment to determine the
development of low-frequency vibrational modes with increasing helix
length for two different terminations. Their geometries are 
fully relaxed PBE+vdW structures. 

For additional details about these calculations we refer the reader to the Supporting Information.

\section{Results and Discussion}

\subsection{Conformational energy hierarchy}

\begin{figure*}[htbp]
\centering
\includegraphics[width=\textwidth]{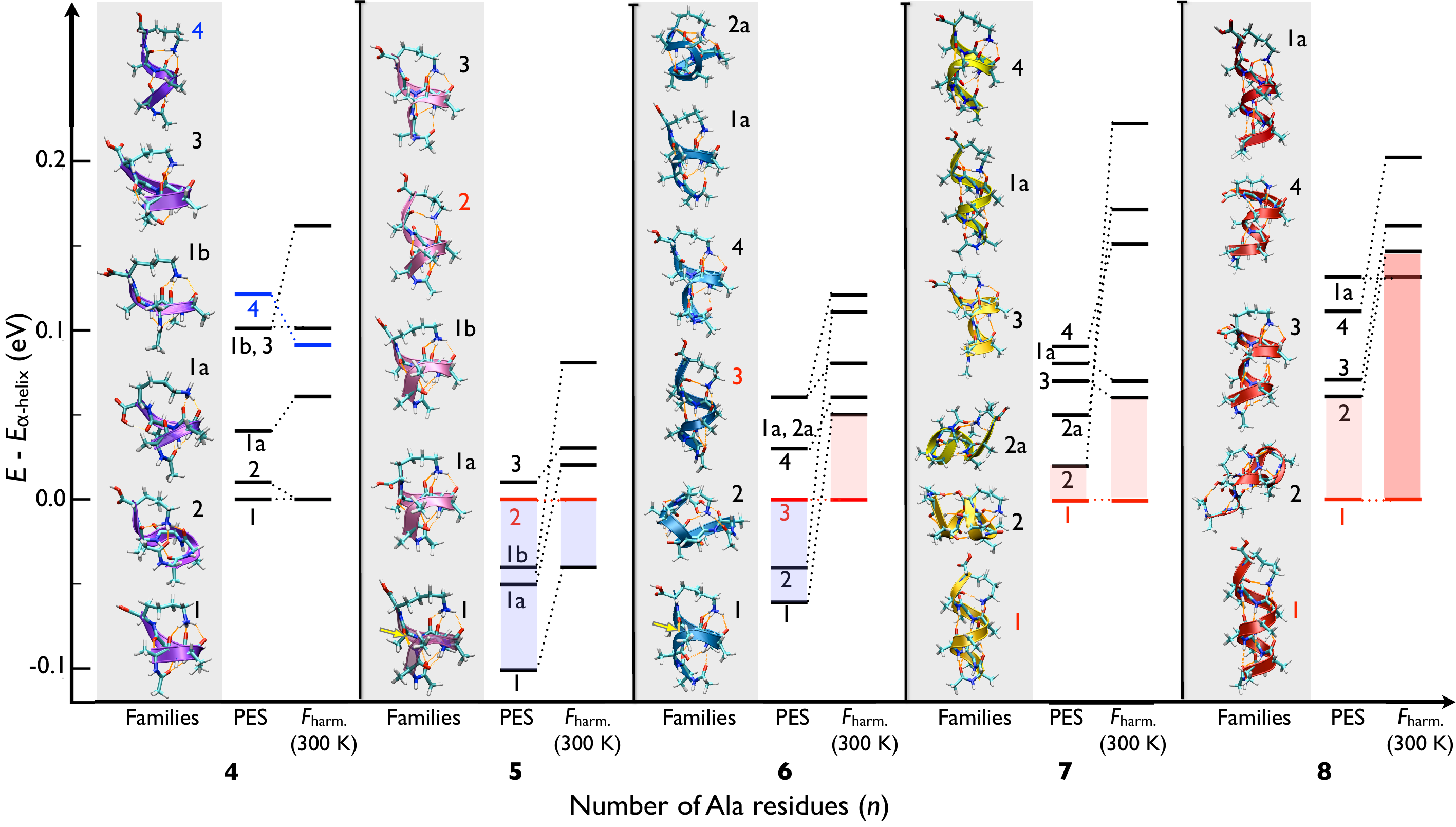}
\caption{Energy hierarchies (thick horizontal bars), obtained with the
  PBE+vdW functional, for the low-energy H-bond families of  
Ac-Ala$_n$-LysH$^+$, $n$=4-8. For each $n$, we include the
conformer representatives of the lowest-energy families up to 0.12 eV
from the global minimum, as defined by local minima of the PBE+vdW 
potential-energy surface (PES). We also show their hierarchy after adding the corrections
for the harmonic vibrational free energy $F_{harm.}$ at $T$=300 K. Numbers labeling each family, 
as well as colored structural representations are shown. The placement
of the structure pictures in the shaded gray areas is not directly
related to the energy axis ($y$ axis), which applies strictly only to
the horizontal bars. For these,
the $\alpha$-helical conformer of $n \geq$5 is chosen as the reference (zero) energy.
For $n$=4, Family 1 is taken as the reference.
$\alpha$-helical conformers are highlighted by red bars.
The 3$_{10}$-helical conformer for $n$=4 is highlighted by blue
bars. The shaded areas in the energy hierarchies indicate the energy
difference between the $\alpha$-helix and the nearest non-helical conformers.
\label{fig:fig1}}
\end{figure*}

Fig. \ref{fig:fig1} summarizes the energetic ordering of the
lowest-energy (PBE+vdW) H-bond families for $n$=4--8, found
employing our search strategy. Only the energy of the
lowest energy structure belonging to each family is reported, and
families are included up to 0.12 eV ($\approx$ 3 kcal/mol) of the lowest identified
minimum of the PBE+vdW PES.
$\alpha$-helical conformers are highlighted in red. 
The 3$_{10}$-helical conformer for $n$=4 is highlighted in blue.
We define purely $\alpha$- (or purely 3$_{10}$-) helical conformers as
those where, counting from the N-terminus, all the backbone CO
groups at residues $i$ are either connected to NH groups at residues $i+4$
(or $i+3$) or to the LysH$^+$ side chain (usually the final three or
four CO groups at the C terminus). 
Coordinates and a more detailed
analysis of all the geometries shown in Fig. \ref{fig:fig1} are 
given in the Supporting Information.

The conformer associated with the lowest-energy PES minimum for $n$=4 is rather small, connecting
almost all its backbone CO groups to the LysH$^+$ termination. The
remaining H-bond at the N-terminus is bifurcated,
comprising an $\alpha$- and a 3$_{10}$-helical H-bond. 
This conformer could therefore be classified as 
the smallest possible $\alpha$-helical prototype in this series.
In contrast, the lowest-energy PES minima for $n$=5 and 6 are
\emph{not} simple helices. Each contains an 
``inverted'' H-bond where one CO group points to the
N-terminus and its connecting NH group points to the C-terminus, producing
more compact structures. In  Fig. \ref{fig:fig1},
these bonds are  highlighted in yellow and pointed to by an arrow.
For $n$=5, we have previously denoted
this conformer as ``g-1'' \cite{Rossi2010}. 
For $n$=7 and 8,
the lowest-energy PES minima correspond to $\alpha$-helices. In each
case, they are closely followed by a conformer that we characterize
as compact/globular (Families 2 of $n$=7 and 8, with an energy
separation of 20 and 60 meV respectively). Thus, we already observe a
cross-over with peptide length to $\alpha$-helical lowest-energy
minima of the PES at $n$=7. 
However, based on the energy hierarchies of the structures of the
local PES minima alone, one would not expect a purely helical
ensemble of conformers at room temperature at $n$=7 or 8. Simple
Boltzmann factors would indicate a mix of structure candidates. Yet, the
experimental work by Kohtani and Jarrold
\cite{KohtaniJarroldWater2004} does indicate a complete
room-temperature structural cross-over at $n$=8 at the latest, albeit
based on a completely different line of reasoning (water adsorption
behavior of ``helical'' versus ``globular'' conformers).

For the low-energy conformers in Fig. \ref{fig:fig1}, we also compute and show in the same figure the
impact of the vibrational free 
energy at room temperature ($T$=300~K) in the harmonic approximation
\footnote{The free energy contributions from
   rigid-body rotations are also included, but 
  differences between these energies for different conformers were
  found to be of the order of only a few meV, with the maximum
  difference amounting to 8 meV.}.
  Remarkably, the relative stability of the
$\alpha$-helices is systematically enhanced by the
vibrational free energy contribution for \emph{all} $n$. 
For $n$=5 and 6 the $\alpha$-helical conformers move down in energy with
respect to the (non-helical) lowest energy PES minimum. For $n$=7 and 8 
the $\alpha$-helices now become the isolated minima. In detail, we observe: \\
- For $n$=8, the energetic interval between the $\alpha$-helical
lowest energy conformer and the nearest globular one (red shaded areas) now amounts to
0.14 eV. The additional Family 1a at 0.13 eV is another $\alpha$-helix
with a slightly modified terminating H-bond network. With the vibrational
free energy included, the energy hierarchy is thus consistent with the
experimental claim that $\alpha$-helices 
dominate over all other possible conformers in gas-phase experiments
for $n$=8 \cite{KohtaniJarroldWater2004, WangLaskin2010}. \\
- For $n$=7, the same qualitative picture emerges. Here, the next
remaining conformer (Family 3) at 300~K is a
mixed 3$_{10}$/$\alpha$-helix. The competing compact conformers (representatives of Family 2 and 2a) are
significantly destabilized by $F_{harm.}$($T$). \\
- For $n$=6, the $\alpha$-helix emerges as the room-temperature
minimum free energy conformer, but the competing non-helical PES
minimum, Family 1, remains close (50 meV). \\
- For $n$=5, the difference between the $\alpha$-helix and Family 1 (g-1) decreases by
60 meV at 300~K, compared to the PES minimum, but Family 1 remains
overall more favorable. \\
In essence, there is an uniform stabilization of helical over more
compact (globular) structures as the temperature increases. The
stabilization tendency increases with peptide length, confirming
quantitatively and systematically from first principles the related
observations in 
Refs. \cite{Kinnear:2002p4458,MaTsaiNussinov2000,
  PlowrightMons2011}. 
We next demonstrate that it is indeed the
\emph{vibrational entropy} term which is critical, and then pinpoint
the physical reason among the low-frequency vibrational modes.

\subsection{Origin of the entropic stabilization}

\begin{figure}[htbp]
\begin{center}
\includegraphics[width=0.6\textwidth]{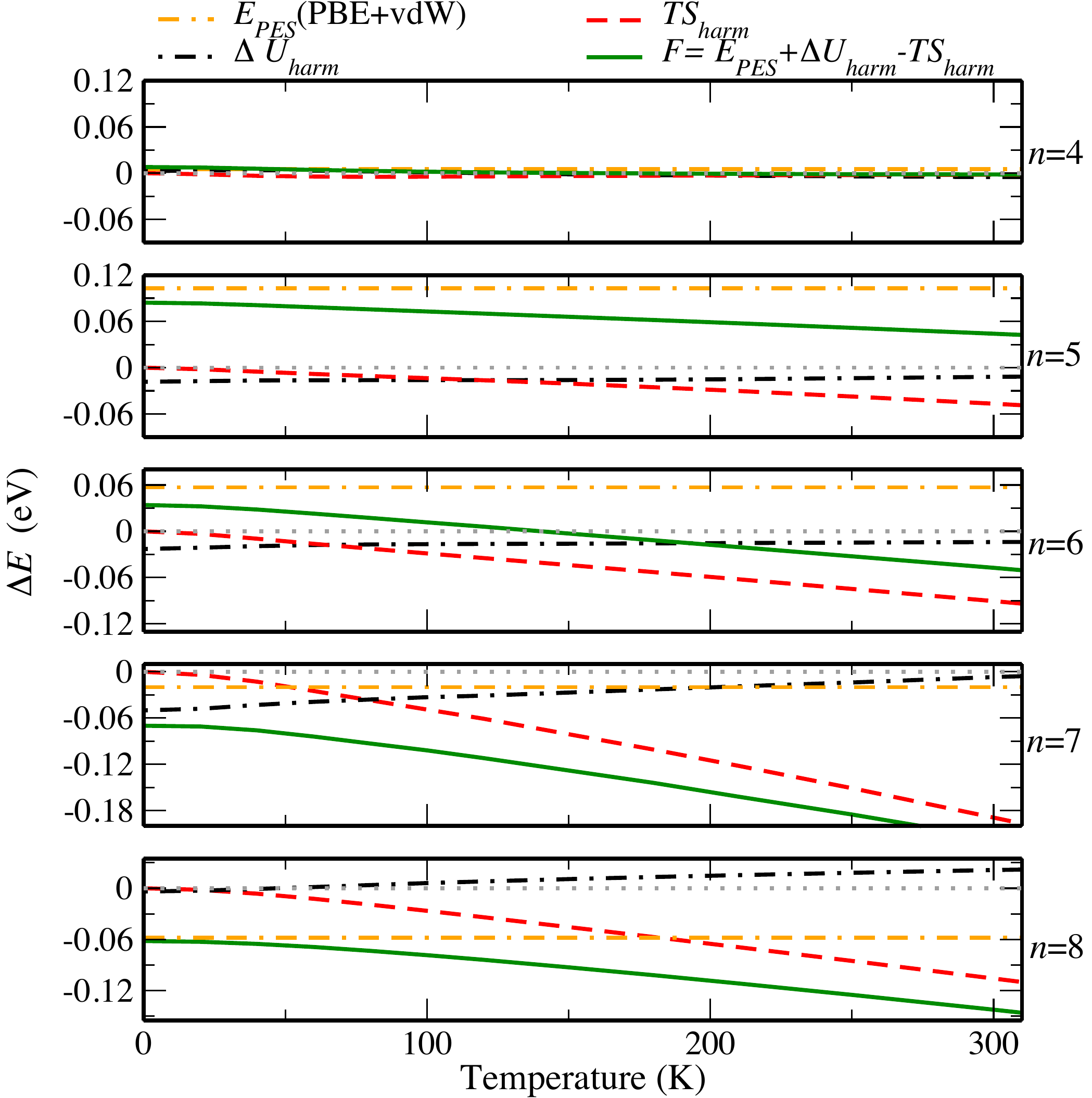}
\caption{Energy differences for each term of the harmonic free energies, as a function of temperature, for: Families 1 and 2 of $n$=4; Families 1 (g-1) and 2 ($\alpha$) of $n$=5; 
Families 1 (``g-1" like) and 3 ($\alpha$) of $n$=6; Families 1 ($\alpha$) and 2 (compact) of $n$=7; and Families 1 ($\alpha$) and 2 (compact) of $n$=8. In each case, the non-$\alpha$-helical conformer was taken as the reference (i.e. negative slopes mean $\alpha$-helix stabilization). 
The PBE+vdW energy differences are plotted with a yellow dash-dash-dot line, the
harmonic internal energy $\Delta U_{harm.}$ as black dash-dot lines, the
entropy term $TS_{harm.}$ as red dashed lines, and the sum $F_{harm.}$ as
green full lines. \label{fig:fig3}}
\end{center}
\end{figure}

In Fig. \ref{fig:fig3} we analyze 
the individual quantities composing the
vibrational free energy differences between: 
Families 1 and 2 of $n$=4, 
Families 1 and the $\alpha$-helical conformers of $n$=5 and 6, as well
as Families 1 ($\alpha$-helical) and 2 for $n$=7 and 8. 
The energy terms plotted are the 
PBE+vdW PES minimum energy, the harmonic internal energy (containing the zero-point energy) $\Delta U_{harm.} (T)$, 
and the entropy term $TS_{harm.}(T)$. 
They are reported in Fig. \ref{fig:fig3} as
energy differences, taking the non-helical conformer of each $n$ as the reference, such that
negative slopes correspond to a stabilization of the $\alpha$-helices. 
Upon inspection of Fig. \ref{fig:fig3}, we
observe a monotonic stabilization of all helical conformers with increasing $T$. 
While for the shortest molecule ($n$=4) there is hardly any observable effect, the stabilization trend is enhanced
with increasing length. For $n$=6,  we predict a 
cross-over of the lowest-energy structures at $T \approx$ 150 K. 
It is clear that among the individual contributions to
the vibrational part of the free energy $F_{harm}(T)$, the entropy term
$TS_{harm}(T)$ is indeed always the most important helix-favoring term. The
zero-point-energy ($\Delta U_{harm}$ at $T$=0), is also favorable, but on
a smaller scale.

The observed stabilization of the $\alpha$-helical conformers can be
understood 
by analyzing the lowest-energy vibrational modes of different conformations.
For simple helices, 
these modes have been estimated and analyzed before \cite{ItohNishi2011}. The important
point here is that we 
have available a direct comparison against many other, non-helical
structure candidates. 
These modes dominate in the harmonic free energy expression at room 
$T$, so that conformers with lower-frequency modes will effectively be
stabilized over others. In Table \ref{tab:all-first-modes}, the
frequencies (in cm$^{-1}$) corresponding to 
the first normal modes of all conformers shown in Fig. \ref{fig:fig1}
are reported. 
The lowest-energy $\alpha$-helices, marked in red in 
Table \ref{tab:all-first-modes}, show first vibrational normal modes 
between 8~cm$^{-1}$ and 13~cm$^{-1}$. The same is true for families 1a
of $n$=7 and 8, which only differ in details of the termination.
In contrast, the conformers that have first vibrational modes of at
least 20~cm$^{-1}$ are all compact non-helical, for example: 
28 cm$^{-1}$ for Family 2 of $n$=7, and 22 cm$^{-1}$ for Family 1 (g-1 motif) of $n$=5.
The g-1 motif has the first vibrational mode lying around 20~cm$^{-1}$
for $n$=5, 6, and 7 (marked with a \textbf{*} symbol in  Table \ref{tab:all-first-modes}).

\begin{figure}[htbp]
\begin{center}
\includegraphics[width=0.5\textwidth]{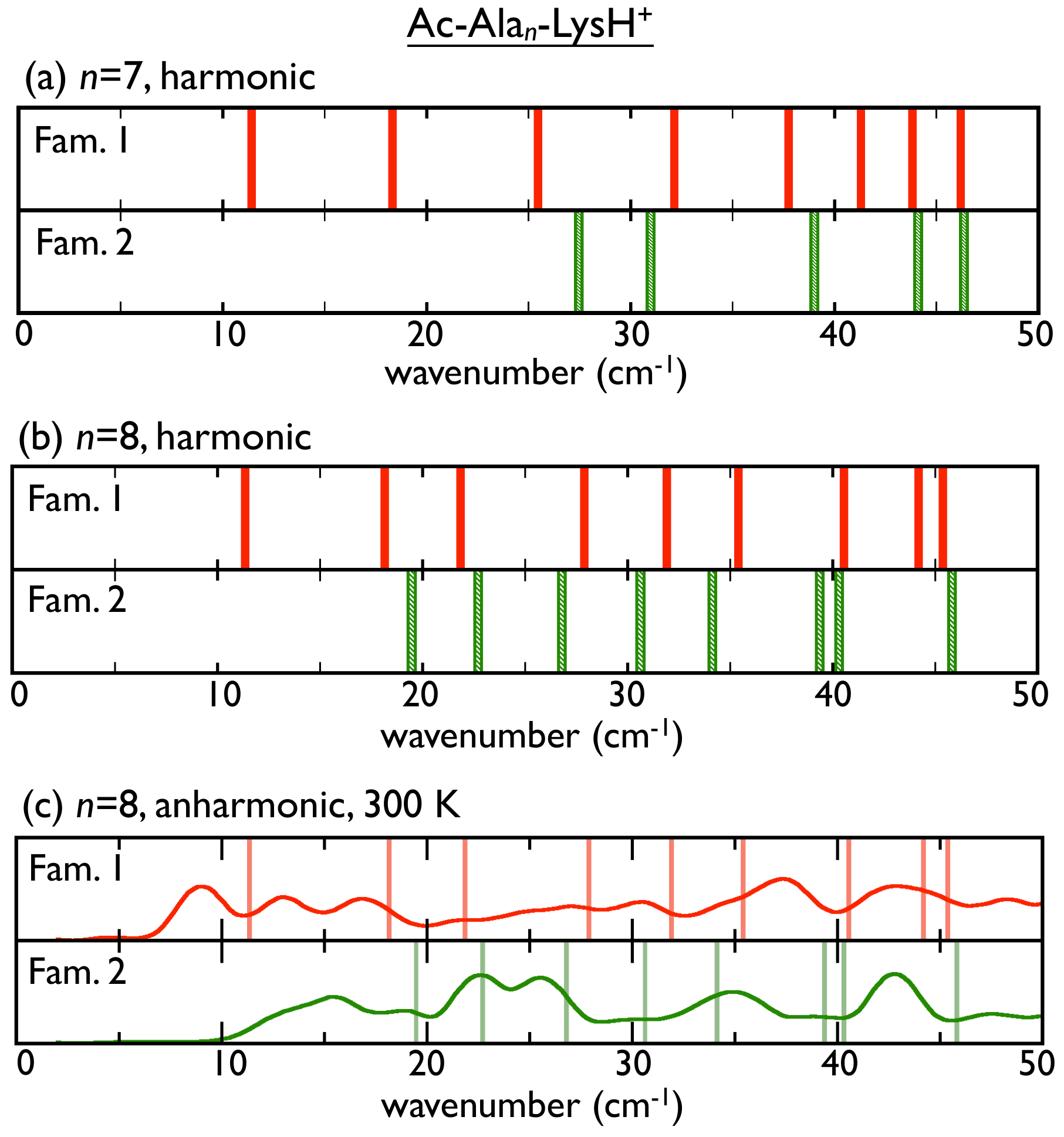}
\caption{Vibrational normal
  modes with frequencies up to 50 cm$^{-1}$ for (a) 
Family 1 ($\alpha$) and Family 2 (globular) of Ac-Ala$_7$-LysH$^+$ and 
(b) Family 1 ($\alpha$) and Family 2 (globular) of Ac-Ala$_8$-LysH$^+$.
(c) Vibrational density of states taken from the velocity autocorrelation function of 21ps of microcanonical AIMD at $\langle T \rangle =$ 300 K, for Family 1 ($\alpha$) and Family 2 (globular) of Ac-Ala$_8$-LysH$^+$. Light colored bars in (c) correspond to the harmonic normal frequencies of vibration [same as (b)]. \label{fig:fig4}}
\end{center}
\end{figure}

Of course, the energy destabilization of the compact conformers with respect
to the helices will depend not only on the first vibrational mode but
also on the overall distribution of low-frequency modes.
To illustrate this point, we show the frequencies
of vibration lying between 0 and 50 cm$^{-1}$ for Families 1 ($\alpha$)
and 2 (globular) of $n$=7 and 8 in Fig. \ref{fig:fig4}(a) and (b). 
The $\alpha$-helices have lower
first vibrational normal modes, \emph{and} have slightly higher (one or two modes)
densities of modes in this region than the globular conformers.

\begin{table}
\begin{center}
\caption{Position of the lowest vibrational mode, in cm$^{-1}$, for the lowest energy
conformers of each family shown in Fig. \ref{fig:fig1} for $n$=4-8.
We use red characters to indicate the  $\alpha$-helical conformers, a \textbf{*} symbol for the g-1-like
conformers of $n$=5 and 6, and blue characters for the 3$_{10}$-helical
conformer. The numerical accuracy of all computed frequencies is 2
cm$^{-1}$ or better (See supplementary material).   \label{tab:all-first-modes}}
\begin{tabular}{@{\vrule height 8pt depth4pt  width0pt} cccccc }
length/conformer & 4 & 5 & 6 & 7 & 8  \,\, \\ 
\hline
Family 1   & 23 & 22\textbf{*} & 20\textbf{*} & \color{red}{12} &\color{red}{11} \,\,\\
Family 1a & 27 & 23 &  21    & \color{red}{10} &\color{red}{11} \,\,\\
Family 1b & 25 & 26 &    &      &        \,\,\\
Family 2   & 17 & \color{red}{13} & 20 & 28 &  20  \,\,\\
Family 2a &      &       &  23 & 22 & \,\, \\
Family 3   & 27\textbf{*} & 20 & \color{red}{8}   & 15 &  16 \,\,\\
Family 4   & \color{blue}{17} &      & 18 & 20\textbf{*} & 15   \,\,\\
   \hline  
  \end{tabular}
\end{center}
\end{table}

Here, a comment regarding structures that are more elongated
than $\alpha$-helices is in order. 
It is customary to compare the stability of $\alpha$-helices with 3$_{10}$ 
helices and $\beta$-sheets or fully extended structures (e.g., in references \cite{IretaGalvan2003, dannenbergccoperativity2, IsmerIretaNeugebauer2008, ImprotaScuseria2001, Topol2001, WuZhao2001, ShiKallenbach2002, bourkubelkakeiderling2002, peneviretashea2008} and many others). 
Following our rationale above, the more extended the structure
is, the softer the low vibrational modes will be. 
This indeed happens for the most extreme case, the fully extended
structure (FES), where we find the first vibrational mode to
lie around only 2 cm$^{-1}$ (calculated for $n$=8 and 15). In fact, 
entropically stabilized 
$\beta$-sheets in \emph{neutral} polyalanine in the gas-phase have
been suggested in  Ref. \cite{DugourdJarrold2005}. 
For 3$_{10}$ helices, in our own structure searches for $n$=4-8, we find the first vibrational mode to lie
always very close to their $\alpha$-helical counterpart. Accordingly,
it is the enthalpic energy difference that favors $\alpha$-helices
specifically over 3$_{10}$
for all molecules studied.
For infinite periodic structures, 
calculation of phonons and vibrational free energies for $\alpha$-, 3$_{10}$-, and $\pi$-helices, and the FES in Ref. \cite{IsmerIretaNeugebauer2008}
corroborate our results. There, all helices are 
destabilized with respect to the FES at 300 K. The 
$\pi$-helix, which is the most compact among the helices studied in Ref. \cite{IsmerIretaNeugebauer2008}, is most destabilized by the vibrational entropy term.

\subsubsection{Beyond the harmonic approximation: \textit{Ab initio} molecular dynamics}

We next show that the observations discussed in the last paragraph 
should hold also in a fully anharmonic picture. Even at relatively low temperature,
where the structure still stays generally constant, we expect
first the inherent anharmonicity of the local, nearly harmonic PES,
then the lowest-barrier transitions between basins (side chain
rotations), and then transitions between locally different
backbone conformations and H-bond
networks \cite{TournierSmith2003,HongSmith2011} to
contribute. Unfortunately, a direct calculation of these terms (e.g., 
by thermodynamic integration) is computationally prohibitive in DFT.
We can, however, use explicit \emph{ab initio} molecular
dynamics simulations to gain some insights. 
In Fig. \ref{fig:fig4}(c), we
compare the Fourier-transformed velocity autocorrelation functions of the
Family 1 (helical) and Family 2 (compact non-helical) conformers of
$n$=8, extracted from explicit
microcanonical \emph{ab initio} molecular dynamics simulations (21~ps
total time, 1~fs time step, initially thermalized to approximately
room temperature). The Fourier transform of the velocity autocorrelation function corresponds
to a vibrational density of states (VDOS). At $T$=0
and for classical 
nuclei, the autocorrelation function should reflect only the harmonic
eigenmodes of Fig. \ref{fig:fig4}(b).
Compared to these modes, 
the onset of the
calculated VDOS at $T$$\approx$300~K is 
noticeably shifted towards lower frequencies for both conformers, but the
onset frequency for the helix is still significantly lower than for
the non-helical structure. Thus, the lower vibrational
frequencies of the helix are also carried over to the full (anharmonic)
motion of the conformers. In addition, the integral over the VDOS up
to 50 cm$^{-1}$ is 4\% larger for the $\alpha$-helical Family 1 than
for the compact Family 2, i.e., the general downshift of frequencies
in this region is also preserved.

Regarding the overall local structure stability during the AIMD simulation,
we observe that the 
H-bond pattern of Family 2 (compact non-helical) stays
essentially the same throughout the entire simulation (see Figure S.1 in the Supporting Information). In
contrast, Family 1 displays local structural fluctuations in the
helical part, occasionally forming short-lived 3$_{10}$-like H-bond
connections. Similar fluctuations also occur in simulations of longer
helices (e.g., Ac-Ala$_{15}$-LysH$^+$ in Ref.~\cite{Rossi2010}). It
seems plausible that the overall greater ``floppiness'' of the
helix compared to the more compact non-helical H-bond network 
adds another favorable entropic contribution at room temperature. 
This direct observation that we make has been guessed as the motive
for the loss in entropy observed in $\alpha$-helical formation, compared to a PPII helical
structure in Ref. \cite{ShiKallenbach2002}. According to our reasoning, the PPII structure,
being less compact than the $\alpha$-helices should indeed show more fluctuations, just like
$\alpha$-helices do if compared to more compact conformers. 

\subsection{The role of the termination}

\begin{figure*}[htbp]
\hspace{-0.3cm}
\includegraphics[width=0.5\textwidth]{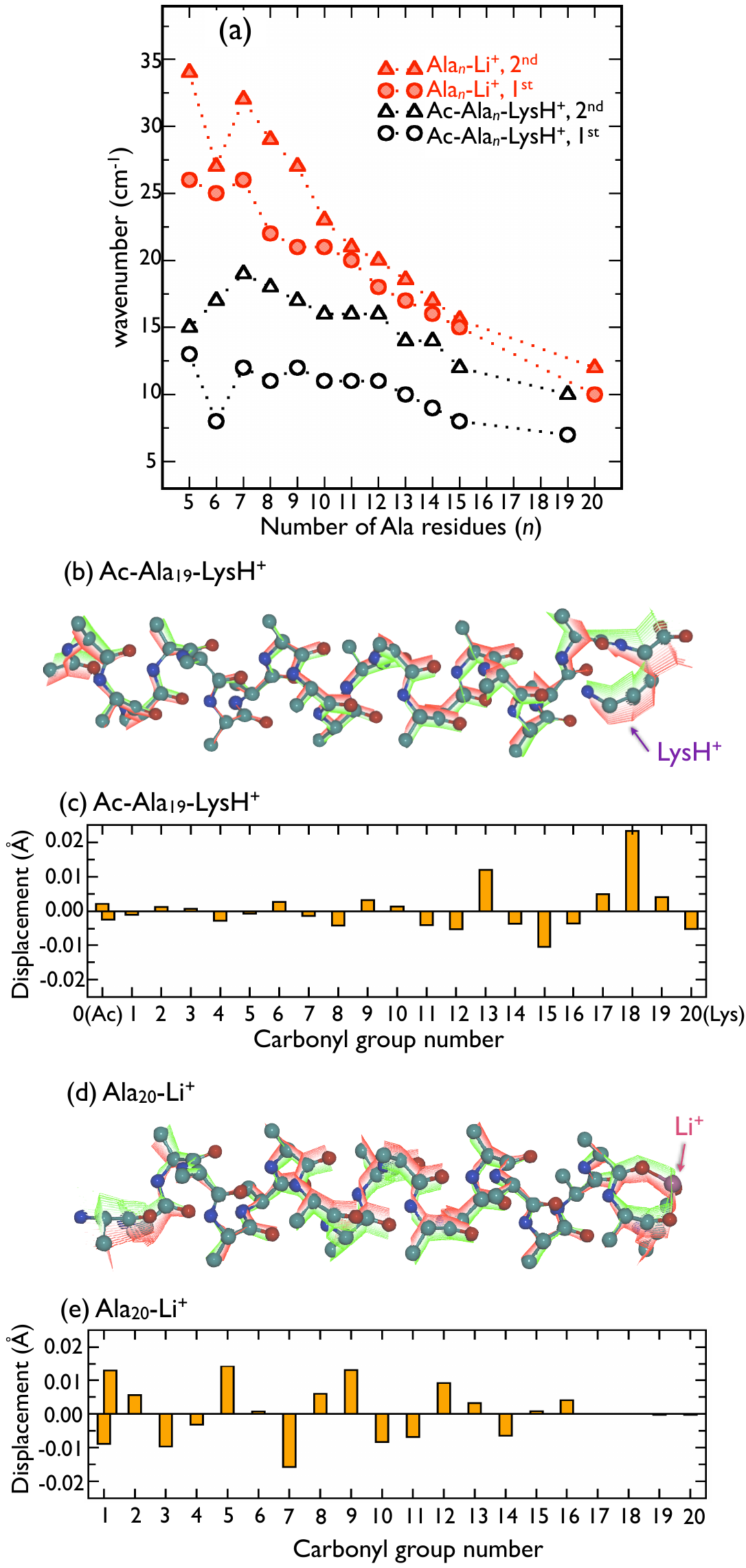}
\caption{Characterization of the lower vibrational modes of helices considered in this work. (a) Positions of the first and
second vibrational modes of $\alpha$-helical
geometries of Ala$_n$-Li$^+$, $n$=5-15, and 20 (red) and $\alpha$-helical
geometries of Ac-Ala$_n$-LysH$^+$, $n$=5-15, and 19 (black). 
The accuracy of these numbers is of around 2 cm$^{-1}$.
(b)  Displacement caused on the backbone atoms of Ac-Ala$_{19}$-LysH$^+$ by
deforming this molecule in the direction of the first vibrational mode, with
red shaded areas corresponding to positive displacements and green shaded areas to negative displacements (hydrogen atoms not shown).
(c) Relative change in H-bond distance when displacing 1 (normalized) unit 
of the first normal mode for Ac-Ala$_{19}$-LysH$^+$.We number the carbonyl groups from the N-terminus to the C-terminus.
(d) Same as (b) for Ala$_{20}$-Li$^+$. 
(e) Same as (c) for Ala$_{20}$-Li$^+$ (in this case the last three carbonyl groups do not make H-bonds, but are connected to the Li$^+$ ion). 
\label{fig:termination_deform}}
\end{figure*}

Finally, we show that, especially for the short peptides considered here,
there are two independent aspects to the stabilization of helical
conformers: length and termination. To isolate their role, we examine
the character of the lowest-frequency modes as a function of
peptide length (also for longer helices) for two different
terminations. The first is the LysH$^+$-terminated series, which is
the main subject of this work. We contrast this series with 
Li$^+$-terminated polyalanine molecules Ala$_n$-Li$^+$, a much more
rigid termination as we shall see. For the latter, we assume
$\alpha$-helices for all $n$ (fully relaxed in DFT-PBE+vdW), which are at least
locally stable when Li$^+$ is placed in contact with the last three 
residues at the C terminus. 
In Fig. \ref{fig:termination_deform}(a), we compare the position of the first and
second vibrational modes of these  $\alpha$-helical
geometries of Ala$_n$-Li$^+$, $n$=5-15, and 20, with the $\alpha$-helical
geometries of Ac-Ala$_n$-LysH$^+$, $n$=5-15, and 19.
For both peptide series, there is a monotonic decrease of
the first and second vibrational frequencies for $n \geq$7, but the starting point
is much higher for the Li$^{+}$ termination (26 cm$^{-1}$ at $n$=7)
than for the LysH$^{+}$ termination. A softening trend of the respective
modes in neutral polyalanine helices with increasing length has also been observed in Ref. \cite{ItohNishi2011}.
In order to characterize  this vibrational mode in more detail, Fig. \ref{fig:termination_deform}(b)
visualizes the displacement of the backbone atoms of Ac-Ala$_{19}$-LysH$^+$ when
deforming this molecule along the first vibrational mode.  Fig. \ref{fig:termination_deform}(c) shows the relative
length changes of the hydrogen bonds in the structure upon deformation
along this mode. Subfigures (d) and (e) show the equivalent data for Ala$_{20}$-Li$^+$.
For both molecules, the vibration spans the helical part of the structures. 
For the LysH$^+$ terminated molecule, we show 
that the actual LysH$^{+}$ termination (connected to the last
four CO residues) is clearly involved in the vibration. In contrast, the
hypothetical Li$^+$ charged termination
constrains especially the C terminus to be much more rigid, as
evidenced by the almost zero change of all Li-O 
distances. Here, the N-terminus is much more involved in the
lowest-frequency modes. The same trends are observed for the smaller
molecules in both series.
Movies containing 3D visualizations of the first vibrational modes for helical
and compact structures
are contained in the Supporting Information.
We thus conclude two points: \\
(i) Helices are entropically favored by allowing delocalized, soft
low-frequency modes that we do not observe in competing, more compact
conformers of the same LysH$^+$ termination (evidenced by 
Table \ref{tab:all-first-modes}). \\
(ii) For short conformers, the already
electrostatically favorable LysH$^+$ termination \cite{PrestaRose1988,
MarquseeBaldwin1989} is additionally helpful by allowing soft,
delocalized modes to include the termination also for short
conformers, in contrast to the hypothetical, much harder
charged termination by Li$^+$. For long enough helices these softer low frequency modes 
should exist regardless of the termination.

\section{Conclusions}

In summary, we show from first principles, quantitatively, and for a
particularly well studied series of polyalanine peptides how
helices emerge with length and temperature as the leading structural
pattern from a vast array of possible competing conformers. The 
crossover to helical stability with length is already apparent based on local structural minima
of the potential-energy surface alone, due to the
critical role of H-bond networks including their
cooperativity \cite{GuoKarplus1994,Baldwin2003,IretaGalvan2003,dannenbergccoperativity2,TkatchenkoRossi2011}
as well as that of vdW terms \cite{TkatchenkoRossi2011}. 
In addition, the contribution from softer low-frequency
vibrational modes acts to stabilize helices with increasing
temperature over their more compact competition. The specific
experimental claim \cite{KohtaniJarroldWater2004} of exclusively
helical conformers at and above $n$=8 is thus explained by \emph{both}
enthalpic and entropic effects acting together at finite temperature. 
\emph{Ab initio} molecular dynamics
simulations corresponding to approximately room temperature suggest
that these trends are further strengthened by anharmonic effects.

The emergence of room-temperature helix stability with length in
Ac-Ala$_n$-LysH$^+$ is thus the result of a subtle balance of
enthalpic and entropic terms. In a non-vacuum environment, further
terms would obviously contribute, but we expect the fact
that helices, in general, allow locally softer vibrational modes to
hold.  Regarding the overall ubiquity of the
helical motif in folded peptides and proteins, at least, we here show
that low-frequency modes will be a significant quantitative
contribution.

\section{Acknowledgement}
The authors would like to acknowledge Dr. Carsten
Baldauf for numerous helpful discussions and suggestions about the
figures and paper in general, including the schematic backbone
vibration visualization in Fig. \ref{fig:termination_deform}.

\section{Supplemental Information}
Additional computational details, XYZ geometries of conformers discussed in this paper, 
detailed information about the structures discussed in this paper, and movies illustrating the first vibrational modes of helices and compact conformers.

\end{document}